\documentclass[twocolumn,amsmath,amssymb]{revtex4} 

\usepackage{graphicx}
\usepackage{dcolumn}
\usepackage{bm}
\usepackage{braket}

\begin{document}

\title{A first-principles model of time-dependent variations in transmission through a fluctuating scattering environment}

\author{Jen-Hao Yeh}
\author{Thomas M. Antonsen}
\author{Edward Ott}
\author{Steven M. Anlage}



\affiliation{Electrical and Computer Engineering Department}
\affiliation{CNAM. Physics Department, University of Maryland,
College Park, MD 20742-4111, USA}

\date{\today}

\begin{abstract}
Fading is the time-dependent variation in transmitted signal
strength through a complex medium, due to interference or temporally
evolving multipath scattering. In this paper we use random matrix
theory (RMT) to establish a first-principles model for fading,
including both universal and non-universal effects. This model
provides a more general understanding of the most common statistical
models (Rayleigh fading and Rice fading) and provides a detailed
physical basis for their parameters. We also report experimental
tests on two ray-chaotic microwave cavities. The results show that
our RMT model agrees with the Rayleigh/Rice models in the high loss
regime, but there are strong deviations in low-loss systems where
the RMT approach describes the data well.

\end{abstract}
\maketitle

Considering wave propagation between a source and a receiver in a
complex medium, fading is the time-dependent variation in the
received signal amplitude as the scattering environment changes and
evolves\cite{Simon}. Fading is a challenging problem in many
situations where waves propagate through a complicated scattering
environment. A common example is the nighttime variation of AM radio
signal reception in the presence of ray bounce(s) off a time
varying ionosphere. Another common observation of fading is
experienced by radio listeners in automobiles moving among vehicles
and buildings in an urban environment. Fading exists in closed or
open scattering systems and in all types of wave propagation, and it
is broadly studied in wireless communication, satellite-to-ground
links, and time-dependent transport in mesoscopic conductors
\cite{Simon,Rayleigh_fading, Foschini,
Delangre,Rice_fading,Lemoine,buttiker}.

The fading amplitude is defined as the ratio of the received signal
to the transmitted signal. The traditional models \cite{Simon} of
fading work well in certain regimes of radiowave propagation
applications, where different probability distribution functions are
chosen depending upon the circumstances. However, these models are
empirically designed for particular scattering environments and
frequency bands, and different (apparently unrelated) fitting
parameters are introduced in different models. For example, the
Rayleigh fading model applies a one-parameter Rayleigh distribution
for the fading amplitude in an environment where there is no
line-of-sight (LOS) path between the transmitter and the receiver,
such as mobile wireless systems in a metropolitan area \cite{Simon,
Rayleigh_fading, Foschini, Delangre}. The Rice fading model, on the
other hand, applies a two-parameter distribution for situations with
a strong LOS path \cite{Simon, Rice_fading, Lemoine}. The detailed
physical origins of these models, and their parameters, are not
clear.

The complexity of the wave propagation environment is advantageous
from the perspective of wave chaos theory because it means that wave
propagation is very sensitive to details, and a statistical
description is most appropriate. For applying wave chaos approaches,
the system should be in the semiclassical limit where the wavelength
is much shorter than the typical size of the scattering system
\cite{Stock}. Researchers have applied random matrix theory (RMT) in
wireless communication \cite{Tulino} and analyzed the information
capacity of fading channels \cite{Moustakas, Morgenshtern, Kumar},
or the scattering matrix ($S$) and the impedance matrix ($Z$) of the
scattering system \cite{Henry, Brouwer, Fyodorov}. Here we directly
apply the random matrix approach to the fading amplitude.

We derive a RMT-based fading model that includes the Rayleigh and
Rice fading models in the high loss regime, but the RMT model also
works well in the limit of low propagation loss. In addition, the
RMT approach combined with a model of non-universal features reveals
the precise physical meanings of the fitting parameters in the
Rayleigh and Rice models.

Considering a $2\times2$ scattering matrix $S$ which describes a
linear relationship between the input and the output voltage waves
on a network, the two ports can be assumed to correspond to the
transmitter and the receiver. The fading amplitude is equivalent to
the magnitude of the scattering matrix element $|S_{21}|$. We start
with a RMT description of the $2\times2$ universal scattering matrix
$s_{rmt}$ in a wave chaotic system \cite{Brouwer}. This description
assumes total ergodicity and does not account for system specific
information (such as the coupling of the ports and the short ray
paths between the ports). For time-reversal invariant (TRI) wave
propagation, statistics of $|s_{rmt,21}|$ can be generated from RMT,
and the only parameter of the distribution $P(|s_{rmt,21}|;\gamma)$
is the dephasing rate $\gamma$ defined in \cite{Brouwer}. Hemmady et
al. \cite{Experimental_tests_our_work} found the relationship
between $\gamma$ and the loss parameter $\alpha$ of the
corresponding closed system as $\gamma = 4\pi\alpha$. The loss
parameter $\alpha$ is the ratio of the closed-cavity mode resonance
3-dB bandwidth to the mean spacing between cavity modes,
$\alpha\equiv f/(2Q\Delta f)$.  Here $f$ is the frequency, $Q$ is the
typical quality factor, and $\Delta f$ is the mean spacing of the
adjacent eigenfrequencies. For an open fading system, we consider an
equivalent closed system in which uniform absorption accounts for
wave energy lost from the system, and we assume that we can define
an equivalent loss parameter $\alpha$ for the open system
\cite{Experimental_tests_our_work}.

We can analytically derive the distribution of the fading amplitude
$P(|s_{rmt,21}|;\alpha)$ for special cases of $\alpha$. For a
lossless system ($\alpha = 0$), the distribution of fading amplitude
is a uniform distribution between 0 and 1. For high loss systems
($\alpha\gg1$), we can prove that the distribution of fading
amplitude goes to a Rayleigh distribution
$P(x=|s_{rmt,21}|;\sigma)=\frac{x}{\sigma^{2}}\exp\left(\frac{-x^{2}}{2\sigma^{2}}\right)$
with the relationship
\begin{equation}\label{alpha_sigma}
\alpha = \frac{1}{8\pi\sigma^{2}}.
\end{equation}
This result reveals the physical meaning of the $\sigma$ parameter
of the Rayleigh fading model, which assumes that the real and
imaginary parts of the complex quantity $S_{21}$ are independent and
identically distributed Gaussian variables with 0 mean and variance
$\sigma^{2}$.

\begin{figure}
\includegraphics[height=1.4in,width=2.0in]{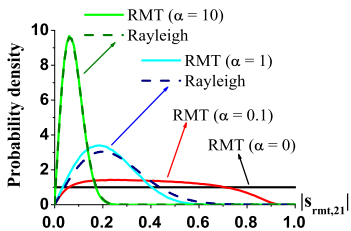}
\caption{The probability density functions $P(|s_{rmt,21}|)$
generated from two theoretical models. Solid curves show the
numerical results from the RMT model with different loss parameters
($\alpha = 0, 0.1, 1, 10$). For higher loss cases ($\alpha = 1$ and
$\alpha = 10$), the corresponding Rayleigh distributions are shown
as dashed curves.}\label{fig_theory}
\end{figure}

In Fig. \ref{fig_theory} we illustrate $P(|s_{rmt,21}|;\alpha)$ for
different loss parameter values from the RMT fading model. For
higher loss cases, we also plot the corresponding Rayleigh
distributions from Eq. (\ref{alpha_sigma}) to show the convergence
of the two models in the high loss limit. Note that distributions
from the RMT model in the low loss region ($\alpha \leq 0.1$)
deviate from a Rayleigh distribution.

To apply wave chaos theory to practical systems, in addition to the
universal $s_{rmt,21}$, we also need to account for the non-chaotic
features of the wave system. We employ the random coupling model
(RCM) \cite{Hart,Jenhao,Experimental_tests_our_work}, which combines
universal fluctuating properties of a scattering system with the
non-universal features arising from the port geometry and
short-orbits between the ports, in the impedance matrix description. The
impedance $(Z)$ matrix specifies the linear relationship between the
port voltages and the port currents, and is related to the
scattering matrix by $Z=Z_{0}(1+S)/(1-S)$, where $Z_{0}$ is a
diagonal matrix with elements equal to the characteristic impedances
of the transmission lines connected to the ports. Another method to
deal with the coupling between the scattering channels and the
system is the Poisson kernel that presents the non-chaotic features
as an average $\bar{S}$ in the scattering matrix description \cite{Gopar,
Kuhl}. The advantage of the RCM is that it can separate the chaotic
(fluctuating) part and the non-chaotic (average) part in a simple
additive format;
\begin{equation}\label{normalized_impedance}
Z^{(the)} =  iX_{avg} + R_{avg}^{1/2}(z_{rmt})R_{avg}^{1/2}.
\end{equation}
$Z^{(the)}$ is the theoretical prediction of the raw measured
impedance matrix $Z$, and $R_{avg}$ and $X_{avg}$ are the real and
imaginary parts of the system-specific ensemble-averaged impedance
matrix $Z_{avg}$. The matrix $Z_{avg}$ can be approximated by taking
the average of the impedances of all realizations in a finite
ensemble, $Z_{avg}=\bar{Z}$. The chaotic part is the universal
impedance matrix $z_{rmt}=(1+s_{rmt})/(1-s_{rmt})$.

In the extended RCM \cite{Hart,Jenhao}, Hart et al. express the
system-specific features in the ensemble wave-scattering system as
\begin{equation}\label{SOC_impedance}
Z_{avg}^{(M)} = iX_{rad} + R_{rad}^{1/2}(z^{(M)}_{so})R_{rad}^{1/2},
\end{equation}
where $R_{rad}$ and $X_{rad}$ are the real and imaginary parts of
the radiation impedance matrix $Z_{rad}$, which is a diagonal matrix
that quantifies the radiation and near-field characteristics of the
ports. The other system-specific feature is short (major) trajectory
orbits. We define an orbit as a ray trajectory that originates from
one port, bounces on the boundary of the system or on scattering
objects, and then reaches a port. Note that the line-of-sight signal
between the two ports is the first (shortest) orbit, and that a
short orbit is distinct from a periodic orbit in a closed system
\cite{periodic_orbit}. We can compute the short-orbits contribution
matrix $z^{(M)}_{so}$ of the $M$ shortest orbits from the known
geometry in each realization of the ensemble \cite{Hart, Jenhao}.
The matrix elements
$z^{(M)}_{so,a,b}=\sum_{m=1}^{M}P_{a,b}^{(m)}C_{a,b}^{(m)}\exp(iS_{a,b}^{(m)})$
is the sum of the short-orbit terms of the $M$ shortest orbits
between port $a$ and port $b$. In each term we consider scattering
on dispersing surfaces (in the geometry factor $C_{a,b}^{(m)}$), the
survival probability ($P_{a,b}^{(m)}$) of the orbits due to the
presence of mobile perturbers, and the propagation phase advance and loss
(in the action $S_{a,b}^{(m)}$). The contribution of trajectory
orbits decreases exponentially with the orbit length
\cite{Hart,Jenhao}. In a lossy system $Z_{avg}^{(M)}\rightarrow
Z_{avg}$ as $M$ increases \cite{Hart, Jenhao}, and only a limited
number of short orbits are required to represent system-specific
features that survive the ensemble average.

According to RMT, the universal complex parameter $s_{rmt,21}$ has
zero mean, but the system-specific features of the sum of short
orbits $z^{(M)}_{so}$ brings about a non-zero bias in the impedance
matrix $Z_{avg}^{(M)}$ (or $Z_{avg}$). Therefore, the measured
$S_{21}$ can have a non-zero mean, and this is similar in character
to the Rice model. The Rice fading model use the distribution
$P(x=|s_{rmt,21}|;\sigma,\nu)=\frac{x}{\sigma^{2}}\exp\left[\frac{-(x^{2}+\nu^{2})}{2\sigma^{2}}\right]I_{0}\left(\frac{x\nu}{\sigma^{2}}\right)$
which contains an additional parameter $\nu$ ($\nu\rightarrow 0$
recovers the Rayleigh distribution), and $I_{0}(\cdot)$ is the
modified Bessel function of the first kind of order zero. The Rice
model is an extension of the Rayleigh model in which the real and
imaginary parts of $S_{21}$ are still independent and identical
Gaussian variables with variance $\sigma^{2}$, but the means are
generalized to a biased mean of magnitude $\nu$. The Rice fading
model is used in environments where one signal path, typically the
line-of-sight signal, is much stronger than the others \cite{Simon,
Rice_fading, Lemoine}, and the $\nu$ parameter is related to the
strength of the strong signal. More generally, we find that the RMT
fading model in the high loss limit yields an explicit expression
for $\nu$ in terms of the short orbit impedance
\begin{equation}\label{nu}
\nu=|s_{so,21}^{(M)}|=\left|\frac{2z_{so,21}^{(M)}}{(1+z_{so,11}^{(M)})(1+z_{so,22}^{(M)})-(z_{so,21}^{(M)})^{2}}\right|.
\end{equation}
This result generalizes the meaning of $\nu$ to include the
influence of all major (short) paths. Note that when there is a
single strong signal that dominates the sum of all paths, the $\nu$
parameter reverts to the original interpretation of Rice fading.

We have carried out experimental tests of the RMT fading model by
measuring the complex $2\times2$ scattering matrix $S$ in two
quasi-two-dimensional ray-chaotic microwave cavities. Both of these
cavities have two coupling ports, which we treat as a transmitter
and a receiver. Microwaves are injected through each port antenna
attached to a coaxial transmission line of characteristic impedance
$Z_{0}=50\Omega$, and each antenna is inserted into the cavity
through a small hole in the lid, similar to previous setups
\cite{Experimental_tests_our_work, Jenhao, Dietz}. The waves
introduced are quasi-two-dimensional for frequencies below the
cutoff frequency for higher order modes ($\sim 19$ GHz) due to the
thin height of the cavities (8 mm in the $z$-direction).

\begin{figure}
\includegraphics[height=1.5in,width=3.4in]{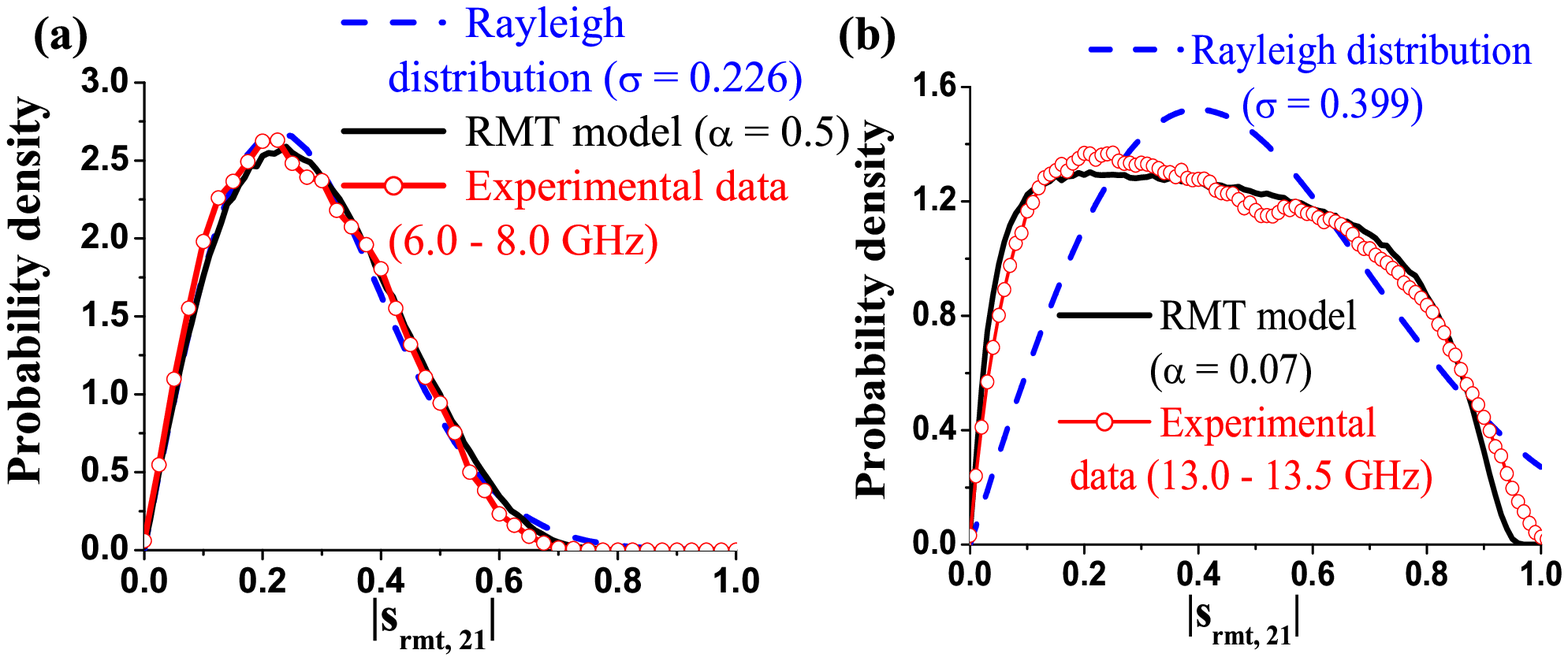}
\includegraphics[height=1.3in,width=1.0in]{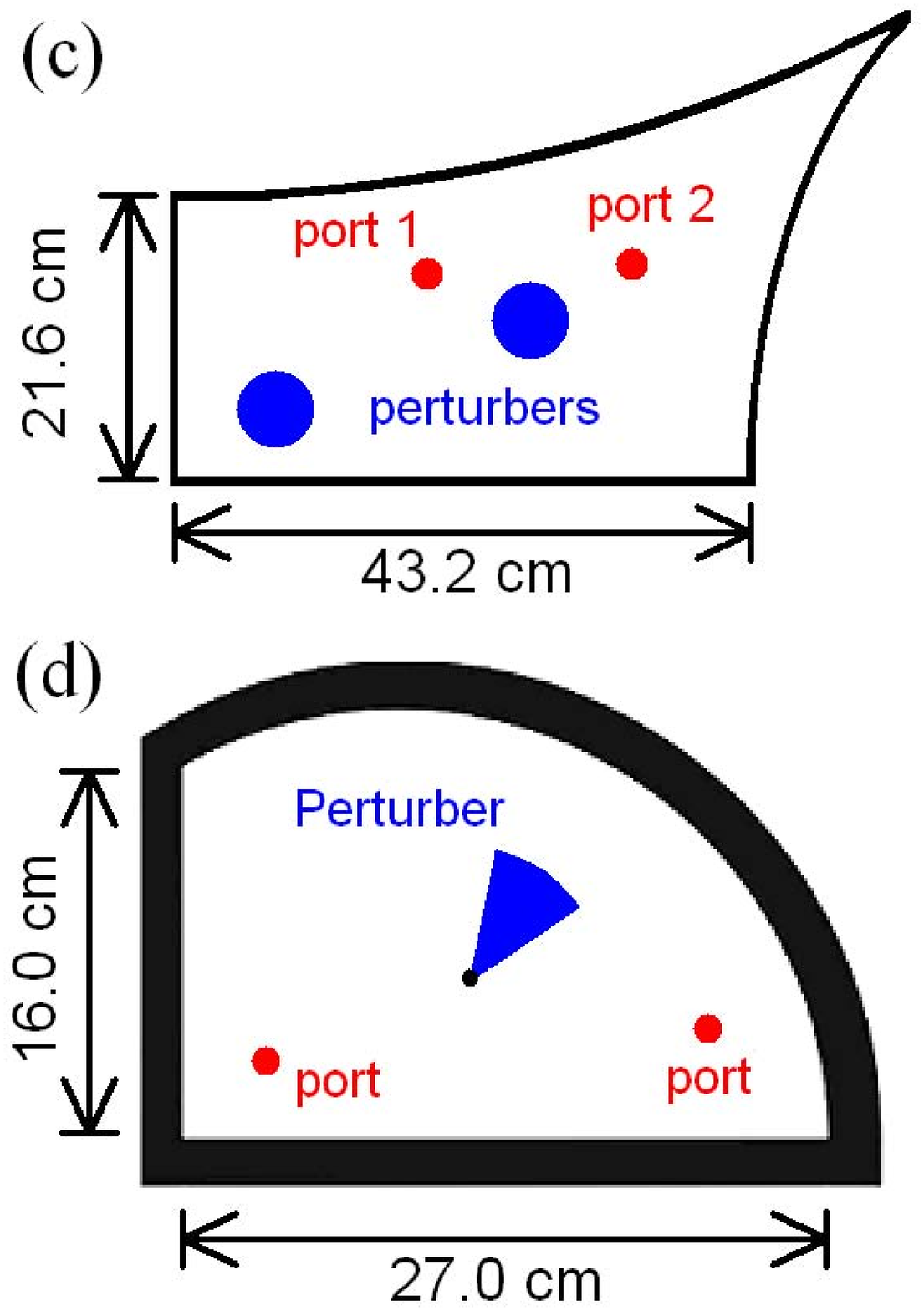}
\includegraphics[height=1.3in,width=1.8in]{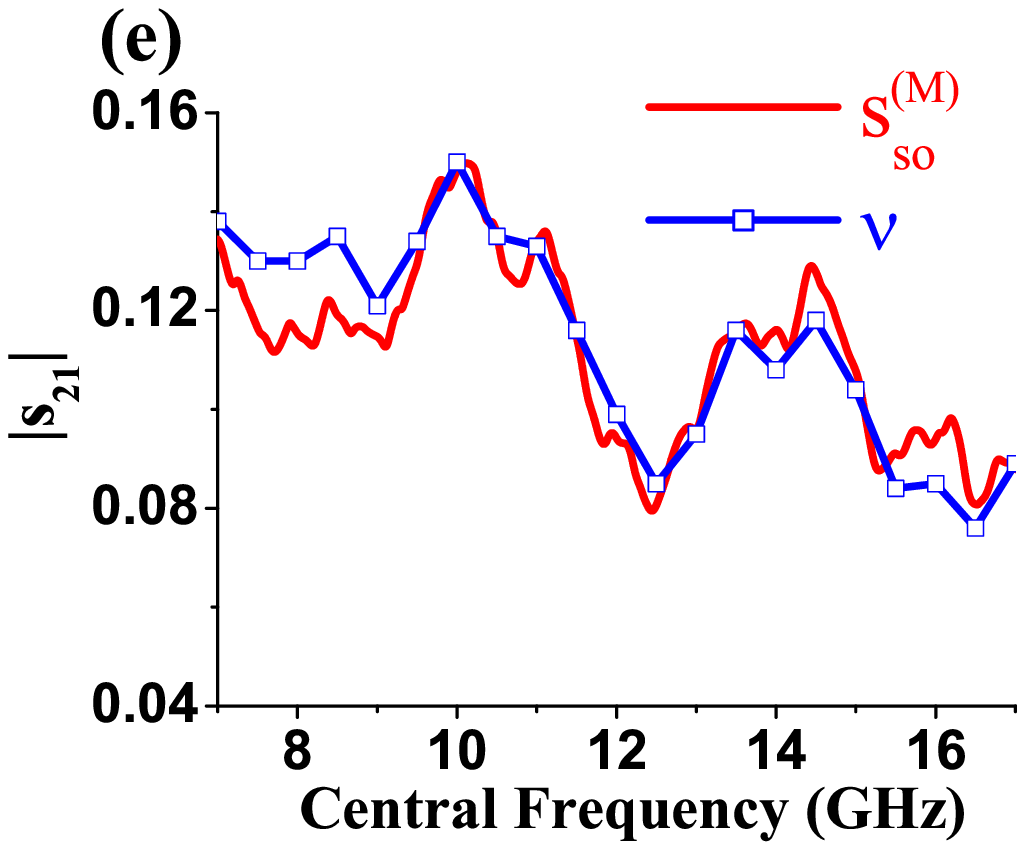}
\caption{Probability density functions $P(|s_{rmt,21}|)$ from the
experimental data (red circles) in (a) the $1/4$ bow-tie cavity and
in (b) the cut-circle cavity, comparing with the RMT model (black
solid) and the best-matched Rayleigh distribution (blue dashed). (c)
The $1/4$ bow-tie cavity with the two ports as red dots and the two
perturbers as blue circles. (d) The cut-circle cavity with the two
ports as red dots and the perturber as a blue wedge. (e) Magnitude
of $|s_{so,21}^{(M)}|$ averaged over a 2-GHz frequency band versus
the central frequency in the $1/4$ bow-tie cavity, and the $\nu$
parameter of the best-matched Rice distribution.
}\label{fig_normalized_s21}
\end{figure}

Classical ray-chaos arises from the shape of the cavity walls. One
cavity is a symmetry-reduced ``bow-tie billiard'' made up of two
straight walls and two circular dispersing walls
\cite{Experimental_tests_our_work, Jenhao} shown in Fig.
\ref{fig_normalized_s21}(c), and the other cavity is a ``cut-circle
billiard'' \cite{Dietz} shown in Fig. \ref{fig_normalized_s21}(d).
The scales of the billiards compared to the wavelengths of the
microwave signals ($1.7 - 5.0$ cm) put these systems into the
semiclassical limit. To create an ensemble for statistical analysis,
we add two cylindrical metal perturbers to the interior of the $1/4$
bow-tie cavity and systematically move the perturbers to create 100
different realizations. For the cut-circle cavity, the perturber is
a Teflon wedge that can be rotated inside the cavity. We rotate the
wedge by 5 degrees each time and create 72 different realizations.
The perturbers can be considered as scattering objects in the
propagation medium, so changing the positions creates the equivalent
of time-dependent scattering variations that give rise to fading.
The $1/4$ bow-tie cavity is made of copper, and measurements of the
transmission spectrum at room temperature suggest the loss parameter goes
from $\alpha=0.4$ to $1.0$, varying with the frequency range
\cite{Jenhao}. The superconducting cut-circle cavity is made of
copper with Pb-plated walls and cooled by a pulsed tube refrigerator
to a temperature of 5.5 K, below the transition temperature of Pb
\cite{Richter, Dietz}. Measurements of the transmission spectrum
suggest $\alpha<10^{-1}$.

The experimental data show good agreement with our RMT fading model.
We first use $Z^{(the)}$ in Eq. (\ref{normalized_impedance}) to
represent the measured impedance matrix $Z$ and solve for $z_{rmt}$
(and therefore $s_{rmt}$). In this process we remove the
system-specific features including all short orbits, so the
situation is equivalent to the Rayleigh fading environment where no
direct paths exist. By choosing data over all realizations in a
frequency range, we can construct the distribution of $|s_{rmt,21}|$
and compare with the prediction of RMT. In Fig.
\ref{fig_normalized_s21} (a) and (b) we plot the distributions of
the fading amplitude from the RMT model (black solid), the
experimental data (red circles), and a best-matched Rayleigh
distribution (blue dashed). In Fig. \ref{fig_normalized_s21}(a) the
room-temperature case, the best-matched RMT model gives a value of
the loss parameter $\alpha = 0.5$ for the experimental data, which
corresponds to $\sigma \simeq (8\pi\alpha)^{-0.5} = 0.282.$ The
best-matched Rayleigh distribution yields $\sigma = 0.226.$ The
difference in $\sigma$ values is due to the fact that the loss
parameter is not very large in this case. Nevertheless, both models
agree with the experimental data well in this loss regime. In Fig.
\ref{fig_normalized_s21}(b) the superconducting cavity case, the
agreement between the experimental data and the RMT model is much
better than the Rayleigh distribution. In fact, in the very-low-loss
region ($\alpha << 1$), the long exponential tail of a Rayleigh
distribution can never match the RMT theoretical distribution that
is limited to $0\leq|s_{rmt,21}|\leq 1$.

In the room-temperature case, since the loss parameter is high
enough, we can compare the relationship between the RMT model and
the Rice fading model [Eq. (\ref{nu})].  In Fig.
\ref{fig_normalized_s21}(e), we compute $z_{so}^{(M)}$ to include
short orbits with length up to 200 cm ($M=1088$) in the $1/4$
bow-tie cavity, apply Eq. (\ref{nu}), perform a sliding average over
a 2-GHz frequency band, and plot the magnitude $|s_{so,21}^{(M)}|$
as the red curve. For the $\nu$ parameter of the Rice model, we
first remove the coupling features from the measured impedance ($Z$)
matrix as $z=R_{rad}^{-1/2}(Z - iX_{rad})R_{rad}^{-1/2}$ and convert
the impedance matrix $z$ to $s$. Then we compare the distribution of
$|s_{21}|$ over a 2-GHz frequency band and 100 realizations with the
best-matched Rice distribution. Since the $\sigma$ parameter has
been determined by the best-matched Rayleigh distribution as
described above for the fully universal data [Fig.
\ref{fig_normalized_s21}(a)], we can use $\nu$ as the only fitting
parameter. In Fig. \ref{fig_normalized_s21}(e) we plot the $\nu$
parameter for the best-matched Rice distributions (blue squares)
along with the system-specific average magnitudes of $s_{21}$ versus
the central frequency of a 2-GHz frequency band. The value of the
Rice $\nu$ parameter and the system-specific feature described by
our model agree well.

One more advantage of applying the RCM is that we can extend the
relations Eq. (\ref{alpha_sigma}) and (\ref{nu}) from the normalized
data to the raw measured data in the high loss cases. In high loss
cases, the magnitude of the elements of $s_{rmt}$ are much less than
one, so we take the approximation to the lowest order
\cite{Gradoni}. For the generalized $\tilde{\nu}$ parameter, we only
need to replace the $z_{so}^{(M)}$ terms by $Z_{avg}$ or
$Z_{avg}^{(M)}$ in Eq. (\ref{nu}). The generalized $\tilde{\sigma}$
parameter is a function of $\alpha$ and all elements of the matrix
$Z_{avg}$. If the transmission between the ports is much less than
the coupling reflection at the ports (i.e.
$|Z_{avg,21}|\ll|Z_{avg,11}|,\ |Z_{avg,22}|$), the modified Rayleigh
$\tilde{\sigma}$ parameter can be simplified to
\begin{equation}\label{general_sigma}
\tilde{\sigma}\simeq\sigma\frac{4\sqrt{Z_{0,11}R_{avg,11}Z_{0,22}R_{avg,22}}}{|Z_{0,11}+Z_{avg,11}||Z_{0,22}+Z_{avg,22}|}.
\end{equation}

In conclusion, we have provided a first-principles derivation of a
RMT fading model that reduces to the traditional Rayleigh and Rice
fading models in high-loss scattering environments, and hence we can
explain the physical meanings of the $\sigma$ parameter of the
Rayleigh distribution and the $\nu$ parameter of the Rice
distribution. Moreover, in low-loss environments, the RMT model can
better predict the distribution of the fading amplitude $|S_{21}|$.
Because wave propagation in a complicated environment is a common
issue in many fields \cite{Simon,Rayleigh_fading, Foschini,
Delangre,Rice_fading,buttiker}, the fading model can be applied to
wireless communication, global positioning systems, and mesoscopic
physics.

We thank the group of A. Richter (Uni.\ Darmstadt) for graciously loaning the
cut-circle billiard. This work is funded by the ONR/Maryland AppEl
Center Task A2 (contract No. N000140911190), the AFOSR under grant
FA95500710049, and Center for Nanophysics and Advanced Materials
(CNAM).


\end{document}